\documentclass[twocolumn,english,superscriptaddress,floatfix]{revtex4}
\usepackage[T1]{fontenc}
\usepackage[latin9]{inputenc}
\usepackage{amsmath}
\usepackage{amssymb}
\usepackage{graphicx}
\usepackage{esint}

\makeatletter
\@ifundefined{textcolor}{}
{%
 \definecolor{BLACK}{gray}{0}
 \definecolor{WHITE}{gray}{1}
 \definecolor{RED}{rgb}{1,0,0}
 \definecolor{GREEN}{rgb}{0,1,0}
 \definecolor{BLUE}{rgb}{0,0,1}
 \definecolor{CYAN}{cmyk}{1,0,0,0}
 \definecolor{MAGENTA}{cmyk}{0,1,0,0}
 \definecolor{YELLOW}{cmyk}{0,0,1,0}
}

\@ifundefined{definecolor}{\usepackage{color}}{}
\@ifundefined{definecolor}{\usepackage{color}}{}
\usepackage{babel}
\usepackage{babel}
\usepackage{babel}
\usepackage{babel}
\usepackage{babel}
\usepackage{babel}
\usepackage{babel}
\usepackage{babel}
\@ifundefined{definecolor}{\usepackage{color}}{}
\usepackage{babel}
\usepackage{babel}

\usepackage{babel}

\usepackage{babel}

\usepackage{babel}

\makeatother

\usepackage{babel}
\begin{document}

\title{Suppression of superconductivity by Neel-type magnetic fluctuations
in the iron pnictides}

\author{Rafael M. Fernandes}

\affiliation{Department of Physics, Columbia University, New York, New York 10027,
USA}

\affiliation{Theoretical Division, Los Alamos National Laboratory, Los Alamos,
NM, 87545, USA}

\author{Andrew J. Millis}

\affiliation{Department of Physics, Columbia University, New York, New York 10027,
USA}

\date{\today }
\begin{abstract}
Motivated by recent experimental detection of Neel-type ($(\pi,\pi)$)
magnetic fluctuations in some iron pnictides, we study the impact
of competing $\left(\pi,\pi\right)$ and $\left(\pi,0\right)$ spin
fluctuations on the superconductivity of these materials. We show
that, counter-intuitively, even short-range, weak Neel fluctuations
strongly suppress the $s^{+-}$ state, with the main effect arising
from a repulsive contribution to the $s^{+-}$ pairing interaction,
complemented by low frequency inelastic scattering. Further increasing
the strength of the Neel fluctuations leads to a low-$T_{c}$ d-wave
state, with a possible intermediate $s+id$ phase. The results suggest
that the absence of superconductivity in a series of hole-doped pnictides
is due to the combination of short-range Neel fluctuations and pair-breaking
impurity scattering, and also that $T_{c}$ of optimally doped pnictides
could be further increased if residual $\left(\pi,\pi\right)$ fluctuations
were reduced. 
\end{abstract}


\maketitle
\begin{figure}
\begin{centering}
\includegraphics[width=0.8\columnwidth]{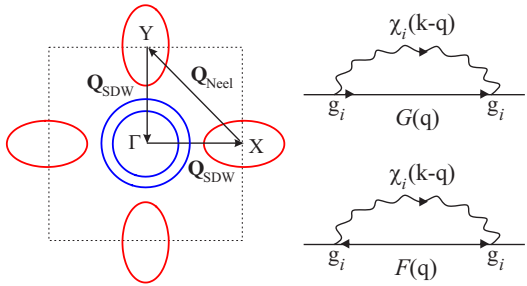} 
\par\end{centering}

\caption{(left panel) Schematic Fermi surface configuration in the 1-Fe Brillouin
zone, with two central hole pockets and two electron pockets. (right
panel) Self-energy diagrams of the Eliashberg equations: normal component
(upper panel) and anomalous component (lower panel).}

\label{fig_FS} 
\end{figure}

The proximity of the superconducting state (SC) to a ``stripe''
spin-density wave instability (SDW) in the phase diagrams of the recently
discovered iron-based superconductors \cite{reviews} (FeSC) prompted
the proposal that SDW spin fluctuations provide the pairing mechanism
\cite{magnetic}. Indeed, the Fermi surface (FS) of many iron pnictides
consists of electron pockets displaced from central hole pockets by
the SDW ordering vector $\mathbf{Q}_{\mathrm{SDW}}=\left(\pi,0\right)/\left(0,\pi\right)$
(see Fig. \ref{fig_FS}). In this situation, even weak SDW fluctuations
may overcome a strong on-site repulsion giving rise to an $s^{+-}$
SC state, in which the gap function has one sign on the electron pockets
and another sign on the hole pockets \cite{reviews_pairing}.

However, the two electron pockets in Fig.~\ref{fig_FS} are connected
by the momentum $\mathbf{Q}_{\mathrm{Neel}}=\left(\pi,\pi\right)$
suggesting that Neel-type magnetic fluctuations may also be important
\cite{Kuroki09}. These fluctuations favor a d-wave SC state in which
the gap function has opposite sign in the two electron pockets. On
the theory side, first-principle and Hartree-Fock calculations find
that the Neel state is locally stable, but with a higher energy than
the SDW state \cite{Johannes09,Bascones11}, while random phase approximation
(RPA) calculations performed in the paramagnetic phase find a peak
in the magnetic susceptibility at $\mathbf{Q}_{\mathrm{Neel}}$, which
is however weaker than the peak at $\mathbf{Q}_{\mathrm{SDW}}$ \cite{Graser10}.

Experimentally, neutron scattering measurements \cite{Mn_neutron}
revealed that even at small $x$, $\mathrm{Ba}\left(\mathrm{Fe_{1-x}Mn_{x}}\right)_{2}\mathrm{As_{2}}$
exhibits spin fluctuations peaked at $\mathbf{Q}_{\mathrm{Neel}}$,
in addition to the SDW fluctuations peaked at $\mathbf{Q}_{\mathrm{SDW}}$.
NMR measurements \cite{Mn_NMR} confirmed that these Neel fluctuations
couple to the conduction electrons. Because the entire family of ``in-plane''
hole-doped $\mathrm{Ba}\left(\mathrm{Fe_{1-x}}M_{\mathrm{x}}\right)_{2}\mathrm{As_{2}}$
compounds ($M=\mathrm{Mn}$, Cr, Mo) \cite{hole_doped_series} displays
SDW order at $x=0$ and Neel order at $x=1$, we expect that competing
Neel and SDW fluctuations will be found across the whole material
family. Intriguingly, superconductivity has not been reported in these
materials to date \cite{Mn_pressure}, in contrast to the electron-doped
counterparts $M=\mathrm{Co}$, Ni, Rh, Pt, Cu, where SC is always
observed \cite{Canfield_transition_metals}.

There is also indirect evidence for Neel fluctuations in the extremely
electron-doped $A_{y}\mathrm{Fe_{2-x}Se_{2}}$ compounds \cite{KFe2Se2}.
In these materials, the near absence of FS pockets in the center of
the Brillouin zone suggests that SDW fluctuations and the $s^{+-}$
state are disfavored, while the square-like shape of the electron
pockets is expected to enhance the $\left(\pi,\pi\right)$ fluctuations
\cite{Scalapino12}. Chemical substitution on the $A$ site or application
of pressure \cite{Balatsky12}, can create a small pocket in the center
of the Brillouin zone, which could support $\left(\pi,0\right)$ fluctuations
and $s^{+-}$ SC.

The effect of competing spin fluctuations on FeSCs is thus of experimental
and theoretical interest. In this paper, we address the problem via
a multi-band Eliashberg approach \cite{Eliashberg,Carbotte} in which
the effect of spin fluctuations on electrons is determined from the
one-loop self energy (see Fig.~\ref{fig_FS}). This approximation
has been extensively employed in studies of cuprates \cite{Monthoux92,Millis92,critical_pairing},
ferromagnetic SC \cite{Roussev01}, and pnictides \cite{eliashberg_pnictides}.
Our calculation goes beyond previous work \cite{eliashberg_pnictides}
by incorporating both SDW and Neel fluctuations, including the Coulomb
pseudo-potential, and using the experimentally determined spin fluctuation
spectrum instead of the single-pole approximation employed previously.

\begin{figure}
\begin{centering}
\includegraphics[width=0.9\columnwidth]{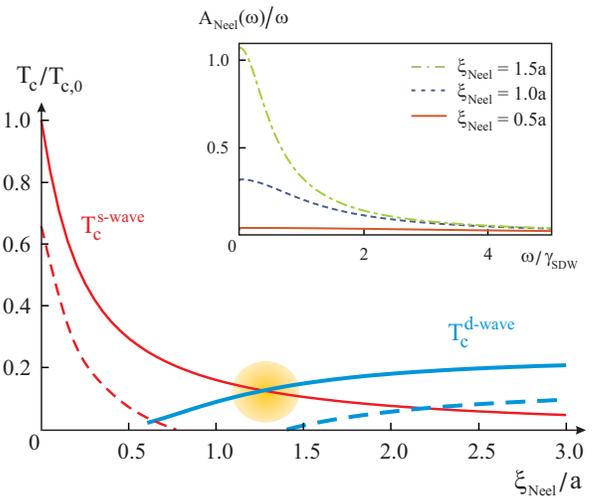} 
\par\end{centering}

\caption{Transition temperatures $T_{c}$ of the s-wave (red/light curve) and
d-wave (blue/heavy curve) states as function of the Neel magnetic
correlation length $\xi_{\mathrm{Neel}}$, for $r=0.65$, $\gamma_{\mathrm{Neel}}/\gamma_{\mathrm{SDW}}=0.33$,
$\lambda_{\mathrm{Neel}}/\lambda_{\mathrm{SDW}}=2$, and $\mu^{*}=0.8$.
$T_{c,0}\approx0.1\gamma_{\mathrm{SDW}}$ is the $s^{+-}$ transition
temperature for $\xi_{\mathrm{Neel}}=0$. The shaded area denotes
the regime where the two states have similar transition temperatures
and a possible $s+id$ state may occur. The dashed lines show the
behavior of the system in the presence of impurity scattering, with\textbf{
$\tau^{-1}\approx0.1T_{c,0}$}. The inset shows the frequency dependence
of the spectral function $A_{\mathrm{Neel}}\left(\omega\right)$ of
the Neel fluctuations for different values of $\xi_{\mathrm{Neel}}$.
\label{fig_phase_diagram}}
\end{figure}

We find that the Coulomb pseudo-potential has only a weak effect on
the dominant $s^{+-}$ state but that even weak, short-range Neel
fluctuations strongly suppress the transition temperature $T_{c}^{\mathrm{s-wave}}$.
If sufficiently strong, the Neel fluctuations may induce a d-wave
state, but the transition temperature is found to be much lower than
the optimal $T_{c}$ for the $s^{+-}$ state. The transition between
$s^{+-}$ and $d$-SC may either occur via an intermediate time reversal
symmetry-breaking $s+id$ state \cite{Stanev10,s_plus_id} or, if
the impurity scattering is stronger, via an intermediate non-SC state
separating the two regions (see Fig. \ref{fig_phase_diagram}).

To gain insight into the results, we use the functional derivative
methods of Bergmann and Rainer \cite{Bergmann_Rainer,Millis88}. We
find that the strong suppression of the $s^{+-}$ state comes mostly
from a repulsive $s^{+-}$ pairing interaction induced by the Neel
fluctuations, although pair-breaking inelastic scattering plays some
role. Finally, we discuss the implications of our results not only
to the SC of the in-plane hole-doped pnictides, but also to the value
of $T_{c}$ in the FeSCs in general.

Our model consists of a two-dimensional FS with two central hole pockets
($\Gamma$, density of states $N_{\Gamma}$) and two electron pockets
($X$ and $Y$, density of states $N_{X}$) displaced from the center
by the momenta $\left(\pi,0\right)$ and $\left(0,\pi\right)$ (Fig.
\ref{fig_FS}) \cite{Maiti11}. For simplicity, hereafter we assume
that these two hole pockets are degenerate - our results do not depend
on this simplification. Following Ref. \cite{Vekhter11}, we set $r=N_{X}/N_{\Gamma}=0.65$.
The electrons are coupled to two types of low-energy bosonic excitations,
namely, SDW spin fluctuations peaked at $\left(\pi,0\right)/\left(0,\pi\right)$
and Neel spin fluctuations peaked at $\left(\pi,\pi\right)$. Experiment
(Refs.\cite{Mn_neutron,INS_pnictides}) indicates that in the paramagnetic
phase these excitations are described by diffusive dynamic susceptibilities:

\begin{equation}
\chi_{i}^{-1}\left(\mathbf{Q}_{i}+\mathbf{q},\Omega_{n}\right)=\xi_{i}^{-2}+q^{2}+\gamma_{i}^{-1}\left|\Omega_{n}\right|\label{susceptib}
\end{equation}

Here, $\mathbf{q}$ is the momentum deviation from the ordering vector
$\mathbf{Q}_{i}$ (all lengths are in units of the lattice parameter
$a$) and $\Omega_{n}$ is the bosonic Matsubara frequency. The quantity
that actually enters the Eliashberg equations is the spectral function
integrated over the momentum component $q_{\parallel}$ parallel to
the FS and evaluated at $q_{\bot}=0$, i.e. $A_{\mathrm{Neel}}\left(\omega\right)=\int dq_{\parallel}\mathrm{Im}\chi_{\mathrm{Neel}}\left(q_{\parallel},\omega\right)$.
This spectral function gives rise to the Matsubara-axis interaction
$a^{(i)}\left(\Omega_{n}\right)=\xi_{i}/\sqrt{1+\left|\Omega_{n}\right|\gamma_{i}^{-1}\xi_{i}^{2}}$
which enters the Eliashberg equations as described below. Note that
the orbital character of the low energy states varies with position
around the FS. In the Eliashberg formalism the resulting angular dependence
of the interaction parameters is averaged over the FS, so as shown
in the Supplementary Material the variation in the orbital character
only affects the values of the effective coupling constants.

The spin fluctuations in each momentum channel $i$ are described
by two parameters: the Landau damping $\gamma_{i}$, which sets the
energy scale, and the correlation length $\xi_{i}$, which sets both
the strength and the spatial/temporal correlations of the spin fluctuations.
We will tune the spectrum by varying $\xi_{i}$. Because the Landau
damping originates from the low-energy decay of the spin excitations
into electron-hole pairs, $\gamma_{i}$ is determined by the electron-boson
coupling constant $g_{i}$ and the densities of states. The coupling
$g_{\mathrm{SDW}}$ is set to yield $T_{c,0}^{\mathrm{s-wave}}\approx30$
K. Following the experimental results of Ref. \cite{Mn_neutron},
we use $\gamma_{\mathrm{Neel}}/\gamma_{\mathrm{SDW}}\approx0.33$
with $\gamma_{\mathrm{SDW}}\approx25$ meV; the value of $g_{\mathrm{Neel}}$
follows from the relationship between $\gamma_{\mathrm{Neel}}/\gamma_{\mathrm{SDW}}$
and $g_{\mathrm{Neel}}/g_{\mathrm{SDW}}$. Finally, we set $\xi_{\mathbf{\mathrm{SDW}}}=5a$
throughout our calculations, varying the correlation length of the
Neel fluctuations $\xi_{\mathrm{Neel}}$. Our results do not change
significantly for smaller values of $\xi_{\mathrm{SDW}}$.

To obtain the transition temperatures in the $s$ and $d$-wave channels
we linearize the Eliashberg equations in the superconducting quantities
and solve the resulting equations for the anomalous component $W_{\alpha,n}$
and the normal component $Z_{\alpha,n}=\mathrm{Im}\Sigma_{\alpha,n}^{N}/i\omega_{n}$
of the self-energy (the real part of $\Sigma^{N}$ just renormalizes
the band dispersions, possibly differently for different pockets \cite{Eliashberg,Carbotte,Benfatto11}).
These quantities are averaged over each Fermi pocket becoming functions
only of the Fermi pocket label $\alpha$ and the fermionic Matsubara
frequency $\omega_{n}=\left(2n+1\right)\pi T$. With the aid of the
auxiliary ``gap functions'' $\bar{\Delta}_{\Gamma,n}\equiv\frac{W_{\Gamma,n}}{Z_{\Gamma,n}\left|\omega_{n}\right|\sqrt{N_{X}}}$
and $\bar{\Delta}_{(X/Y),n}\equiv\frac{W_{(X/Y),n}}{Z_{X,n}\left|\omega_{n}\right|\sqrt{N_{\Gamma}}}$,
the linearized gap equation is expressed as a matrix equation in Matsubara
(indices $n,m$) and band (indices $\alpha,\beta$) spaces $\sum_{m,\beta}K_{nm}^{\alpha\beta}\bar{\Delta}_{\beta,m}=0$,
with the kernel:

\begin{eqnarray}
\left(K_{nm}^{\alpha\beta}\right) & = & -\left(\begin{array}{ccc}
\delta_{nm}\frac{Z_{\Gamma,n}\left|\omega_{n}\right|}{T} & \frac{\lambda_{\mathrm{SDW}}}{2}a_{nm}^{(1)} & \frac{\lambda_{\mathrm{SDW}}}{2}a_{nm}^{(1)}\\
\lambda_{\mathrm{SDW}}a_{nm}^{(1)} & \delta_{nm}\frac{Z_{X,n}\left|\omega_{n}\right|}{T} & \lambda_{\mathrm{Neel}}a_{nm}^{(2)}\\
\lambda_{\mathrm{SDW}}a_{nm}^{(1)} & \lambda_{\mathrm{Neel}}a_{nm}^{(2)} & \delta_{nm}\frac{Z_{X,n}\left|\omega_{n}\right|}{T}
\end{array}\right)\nonumber \\
 &  & -\frac{1}{2}\left[\frac{\mu^{*}}{1+r}-\frac{\tau^{-1}}{T}\frac{\delta_{nm}}{r}\right]\left(\begin{array}{ccc}
2 & \sqrt{r} & \sqrt{r}\\
2\sqrt{r} & r & r\\
2\sqrt{r} & r & r
\end{array}\right)\label{kernel}
\end{eqnarray}

Here we have introduced the matrix elements coming from the bosonic
modes $a_{nm}^{(i)}=a^{i}(\omega_{n}-\omega_{m})$ ($i=1$ corresponds
to SDW and $i=2$, to Neel fluctuations) and the dimensionless coupling
constants $\lambda_{\mathrm{SDW}}=2g_{\mathrm{SDW}}^{2}\sqrt{N_{\Gamma}N_{X}}$,
$\lambda_{\mathrm{Neel}}=g_{\mathrm{Neel}}^{2}N_{X}$. $T$ is the
temperature. We also introduce an upper frequency cutoff $\Lambda=8\gamma_{\mathrm{SDW}}$,
corresponding to the energy scale of the bottom/top of the electron/hole
bands, and we assume that $\mu^{*}$ is a bare Coulomb interaction
renormalized in the standard way by higher energy processes. $\tau^{-1}$
is the scattering rate associated with non-magnetic point impurities
and the $Z_{\alpha,n}$ functions are obtained analytically (see Supplementary
Material). The Coulomb pseudo-potential favors solutions with $\sum_{\alpha}N_{\alpha}\Delta_{\alpha}=0$.

Reflecting the tetragonal symmetry of the system, the matrix equation
supports two different types of solution: the s-wave state $\bar{\Delta}_{X,n}=\bar{\Delta}_{Y,n}$,
with either $s^{++}$ ($\bar{\Delta}_{\Gamma,n}\propto\bar{\Delta}_{X,n}$)
or $s^{+-}$ ($\bar{\Delta}_{\Gamma,n}\propto-\bar{\Delta}_{X,n}$)
structure, and the d-wave state $\bar{\Delta}_{X,n}=-\bar{\Delta}_{Y,n}$.
The solution in a given symmetry channel is obtained when the largest
eigenvalue $\eta$ of the matrix (\ref{kernel}) vanishes. Since our
calculations never yield an $s^{++}$ state, we use the terms s-wave
and $s^{+-}$ to refer to the same state. Due to limitations of the
size of the matrices that can be diagonalized, and since the matrix
size scales as $\Lambda/T$, we resolve $T_{c}\gtrsim10^{-3}\gamma_{\mathrm{SDW}}$.
Hereafter, we set $\lambda_{\mathrm{SDW}}=0.4$ and the Coulomb pseudo-potential
$\mu^{*}=0.8$, which gives, in the absence of competing Neel fluctuations,
$T_{c,0}^{\mathrm{s-wave}}\approx30$K and implies $\lambda_{\mathrm{Neel}}=0.8$.

Fig. \ref{fig_phase_diagram} shows our principal results: the dependence
of the SC transition temperature $T_{c}$ on the strength of Neel
fluctuations (parametrized by the Neel correlation length $\xi_{\mathrm{Neel}}$).
The light solid line (red online) shows the transition temperature
$T_{c}^{\mathrm{s-wave}}$ for the $s^{+-}$ channel in the absence
of impurity scattering. Surprisingly, even weak, short-range fluctuations
strongly suppress $s^{+-}$ SC, but once $T_{c}^{\mathrm{s-wave}}$
has been substantially reduced, the additional suppression effect
caused by further increasing $\xi_{\mathrm{Neel}}$ is small. Sufficiently
strong Neel correlations produce a d-wave solution (heavier solid
line, blue online) with $T_{c}^{\mathrm{d-wave}}$ that eventually
becomes larger than $T_{c}^{\mathrm{s-wave}}$ but always remains
small compared to the maximum $T_{c}^{\mathrm{s-wave}}$. In our linearized
theory the transition between s-wave and d-wave superconductors appears
as a discontinuous change in the nature of the state, but the considerations
of \cite{Stanev10} suggest that nonlinear terms not included here
will generate an intermediate $s+id$ state (shaded area). The dashed
lines show the behavior in the presence of impurity scattering, which
is pair-breaking for both $s^{+-}$ and $d$-wave superconductivity.
Sufficiently strong impurity scattering can disconnect the two SC
states, leaving an intermediate non-SC regime.

\begin{figure}
\begin{centering}
\includegraphics[width=0.9\columnwidth]{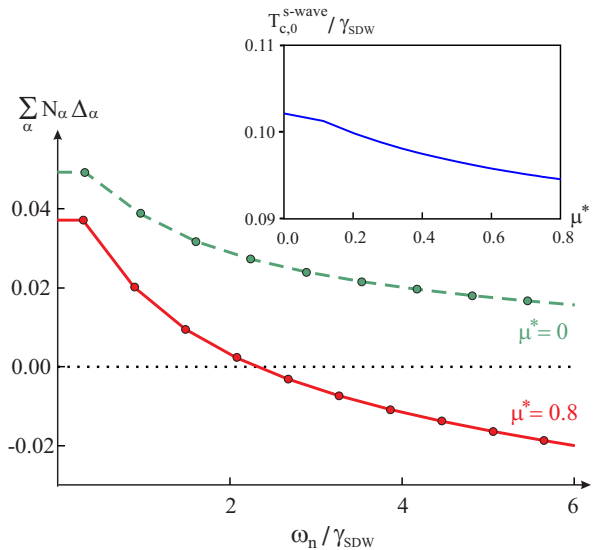} 
\par\end{centering}

\caption{Averaged $s^{+-}$ gap function $\sum_{\alpha}N_{\alpha}\Delta_{\alpha}$
across the different pockets at $T_{c,0}^{\mathrm{s-wave}}$ as function
of Matsubara frequency $\omega_{n}$ (in units of $\gamma_{\mathrm{SDW}}$),
for $\mu^{*}=0$ (green/dashed curve) and $\mu^{*}=0.8$ (red/solid
curve). The inset shows $T_{c,0}^{\mathrm{s-wave}}$ (in units of
$\gamma_{\mathrm{SDW}}$) as function of $\mu^{*}$. Although here
we used $\xi_{\mathrm{Neel}}=0$, a similar behavior holds for $\xi_{\mathrm{Neel}}\neq0$.
\label{fig_pseudopotential}}
\end{figure}

We also analyze the impact of the Coulomb pseudo-potential $\mu^{*}$
on the $s^{+-}$ state - the d-wave state avoids the Coulomb repulsion.
Fig. \ref{fig_pseudopotential} shows the pocket-averaged $s^{+-}$
gap function $\sum_{\alpha}N_{\alpha}\Delta_{\alpha}$ both in the
presence and in the absence of $\mu^{*}$. To avoid the local repulsion,
the averaged order parameter changes sign at a non-zero Matsubara
frequency, although the sign of each individual gap does not necessarily
change. This is the multi-band analogue of the response of a single-band
s-wave superconductor to the local repulsion. For all values of $\xi_{\mathrm{Neel}}$
we have studied, neither $\Delta_{n=0}$ nor $T_{c}^{\mathrm{s-wave}}$
(shown in the inset) are substantially altered by $\mu^{*}$, in agreement
with the weak-coupling analysis of Ref. \cite{Mazin_Schmalian}.

\begin{figure}
\begin{centering}
\includegraphics[width=0.9\columnwidth]{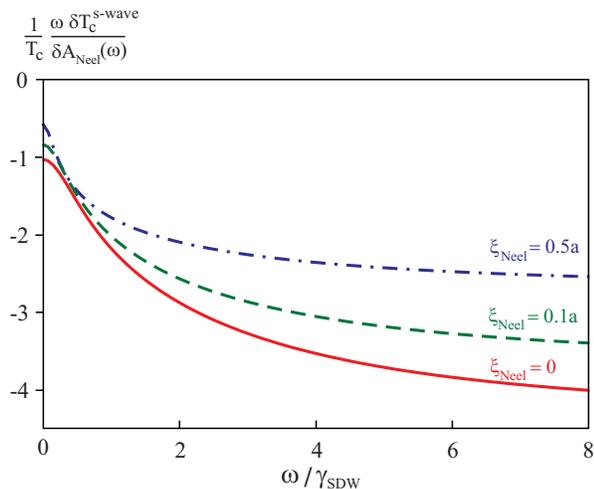} 
\par\end{centering}

\caption{Functional derivative $T_{c}^{-1}\omega\delta T_{c}^{\mathrm{s-wave}}/\delta A_{\mathrm{Neel}}\left(\omega\right)$
as function of frequency $\omega$ (in units of $\gamma_{\mathrm{SDW}}$)
for the cases $\xi_{\mathrm{Neel}}=0$ (red/solid line), $\xi_{\mathrm{Neel}}=0.1a$
(green/dashed line), and $\xi_{\mathrm{Neel}}=0.5a$ (blue/dotted
dashed line). \label{fig_bergmann}}
\end{figure}

We now turn to the physics of the decrease of $T_{c}^{\mathrm{s-wave}}$
caused by Neel fluctuations. Increasing $\xi_{\mathrm{Neel}}$ increases
the spin fluctuation intensity and changes its functional form (see
the inset of Fig.~\ref{fig_phase_diagram}). To analyze how different
frequency regions of the Neel spectral function $\,$affect $T_{c}^{\mathrm{s-wave}}$,
we follow Ref. \cite{Bergmann_Rainer,Millis88} and calculate the
functional derivative 
\begin{equation}
\frac{1}{T_{c}}\frac{\omega\,\delta T_{c}}{\delta A_{\mathrm{Neel}}\left(\omega\right)}=\left[\hat{\Delta}\frac{\omega\,\delta\hat{K}}{T_{c}\,\delta A_{\mathrm{Neel}}\left(\omega\right)}\hat{\Delta}\right]/\left(-\frac{\partial\eta}{\partial T_{c}}\right)_{\eta=0}\label{derivtc}
\end{equation}
for different values of $\xi_{\mathrm{Neel}}$, as shown in Fig. \ref{fig_bergmann}.
Previous work \cite{Bergmann_Rainer,Millis88} has shown that the
low frequency regime captures the pair-breaking effects of inelastic
scattering while the high frequency regime expresses the changes to
the pairing interaction. The larger magnitude of $\delta\log T_{c}/\delta A_{\mathrm{Neel}}$
at high frequencies shows that in the pnictides the dominant effect
of the Neel fluctuations is to provide a negative contribution to
the $s^{+-}$ pairing interaction, with the extra pair-breaking effect
of the induced low frequency inelastic scattering being less important.
Because Fig.~\ref{fig_bergmann} shows that the logarithmic derivative
of $T_{c}$ is a slow function of $\xi_{\mathrm{Neel}}$ we conclude
that the initial steep drop and subsequent flattening of the $T_{c}$
curve shown in Fig.~\ref{fig_phase_diagram} is due in large part
to the variation of $T_{c}$ itself. Additionally, as $\xi_{\mathrm{Neel}}$
is increased the Neel fluctuation spectrum shifts to lower frequencies
(see the inset of Fig.~\ref{fig_phase_diagram}), where the pair-breaking
is less effective. However, additional physics is also at play. In
the weak-coupling limit of two effective competing pairing interactions
$\lambda_{s}$ and $\lambda_{d}$ we obtain $T_{c}\propto\exp\left[-1/\left(\sqrt{\lambda_{s}^{2}+\lambda_{d}^{2}}-\lambda_{d}\right)\right]$
so 
\begin{equation}
\frac{d\log T_{c}}{d\lambda_{d}}\propto-\frac{1}{\left(\lambda_{s}^{2}+\lambda_{d}^{2}-\lambda_{d}\sqrt{\lambda_{s}^{2}+\lambda_{d}^{2}}\right)}\label{weakcoupling}
\end{equation}
which is larger in magnitude for larger $\lambda_{d}$, implying an
opposite ordering of the curves to that seen in Fig.~\ref{fig_bergmann}.
Our Eliashberg results and Eq.~\ref{weakcoupling} differ because
the gap function self-consistently adjusts to the pairing potential,
so that for larger $\xi_{\mathrm{Neel}}$ the gap function decreases
more rapidly with frequency, thereby minimizing the depairing effects
of the Neel fluctuations (see Supplementary Material).

Our results offer a possible explanation for the puzzling behavior
of the hole-doped $\mathrm{Ba}\left(\mathrm{Fe_{1-x}}M_{\mathrm{x}}\right)_{2}\mathrm{As_{2}}$
series ($M=\mathrm{Mn}$, Cr, Mo) \cite{hole_doped_series}, which,
in contrast to its electron-doped counterpart ($M=\mathrm{Co}$, Ni,
Rh, Pt, Cu) \cite{Canfield_transition_metals}, does not display SC.
The short-range Neel fluctuations induced by the dopants, which were
observed experimentally for low concentrations of $M=\mathrm{Mn}$
\cite{Mn_neutron,Mn_NMR}, suppress the $s^{+-}$ state without giving
rise to a high-temperature d-wave state. This low-$T_{c}$ $s^{+-}$
state, in turn, can be easily suppressed, for example by impurity
scattering or by another competing ordered state, such as the SDW
state \cite{FernandesPRB10,Vorontsov09} observed at low $x$. We
suggest that improving the purity of the samples and applying pressure
to suppress the SDW state may reveal either a weakened $s^{+-}$ state
or perhaps a low $T_{c}$ d-wave state. Similarly, in the extremely
electron-doped $A_{y}\mathrm{Fe_{2-x}Se_{2}}$ systems \cite{KFe2Se2,ARPES_Fe2Se2,Keimer_Fe2Se2},
where for small $y$ the hole pocket is generally absent and d-wave
superconductivity is discussed, adding holes by changing $A$ \cite{Scalapino12}
or by applying pressure \cite{Balatsky12} should produce the reverse
competition.\textbf{ }Interestingly, recent pressure experiments found
two separate SC domes in \textbf{$\mathrm{K}_{0.8}\mathrm{Fe_{1.78}Se_{2}}$}
\cite{Mao_FeSe}, which could be related to the behavior shown in
Fig. \ref{fig_phase_diagram} (dashed lines). Indeed, pressure changes
the shapes of the Fermi pockets, which affects the relative strength
of SDW and Neel fluctuations.

More generally, since most FeSC compounds have two matching electron
pockets separated by $\mathbf{Q}_{\mathrm{Neel}}$ even at optimal
doping compositions\textbf{,} we expect at least weak Neel-type fluctuations.
Indeed, recent Raman data indicate that a d-wave instability, presumably
originated from these Neel fluctuations, compete with the $s^{+-}$
state of optimally-doped FeSC \cite{raman_mode}. However\textbf{,
}our findings show that these weak\textbf{ }$\left(\pi,\pi\right)$
fluctuations\textbf{ }strongly suppress\textbf{ $T_{c}^{\mathrm{s-wave}}$}.\textbf{
}This suggests that the highest $T_{c}$ in several FeSCs can be potentially
enhanced if the $\left(\pi,\pi\right)$ fluctuations are minimized.\textbf{
}One possible route is to make the sizes and shapes of the two electron
pockets unequal, via, for example, a tetragonal symmetry breaking
\cite{Fernandes12}. Interestingly, torque magnetometry measurements
found such a tetragonal symmetry breaking above $T_{c}$ in some optimally
doped FeSCs \cite{Matsuda12}. In Ref. \cite{Kuo12}, it was also
observed that a small strain applied along the orthorhombic axis can
enhance $T_{c}$.

In summary, our results open a new route to explore unconventional
superconductivity in multi-band systems by controlling competing spin
fluctuations. In particular, Neel fluctuations have a strong effect
on the $s^{+-}$ state of the FeSCs, rapidly reducing $T_{c}$ and
potentially driving a transition from s-wave to d-wave SC (Fig. \ref{fig_phase_diagram}).
Depending on the strength of impurity scattering, more exotic states
can emerge, such as the $s+id$ state \cite{Stanev10}, although this
might also arise from other mechanisms \cite{s_plus_id}. Notice also
that the lower $T_{c}$ solution, even if not present in the ground
state, will give rise to a collective excitation which can in principle
be detected by Raman scattering \cite{raman_mode,collective_mode}.

We thank P. Canfield, A. Chubukov, A. Goldman, I. Eremin, A. Kreyssig,
B. Lau, R. McQueeney, D. Pratt, J. Schmalian, and G. Tucker for useful
discussions. RMF is supported by the NSF Partnerships for International
Research and Education (PIRE) program OISE-0968226 and AJM by NSF-DMR-1006282.

\begin{widetext}

\vspace{8 mm}

{\bf\Large \center Supplementary material for ``Suppression of superconductivity by
Neel-type magnetic fluctuations in the iron pnictides''} \\

\setcounter{equation}{0}
\renewcommand{\theequation}{S\arabic{equation}}

\setcounter{figure}{0}
\renewcommand{\thefigure}{S\arabic{figure}}

\section{Formulation and Solution of Eliashberg Equations}

\subsection{Formulation of Equations}

The low-energy action describing the coupling between the electrons
and the SDW and Neel fluctuations is conveniently expressed in terms
of the Nambu operator $\Psi_{\mathbf{k}}^{\dagger}=\left(\begin{array}{cccccccc}
c_{\Gamma,\mathbf{k}\uparrow}^{\dagger} & c_{\Gamma,-\mathbf{k}\downarrow}^{\phantom{}} & c_{\Gamma',\mathbf{k}\uparrow}^{\dagger} & c_{\Gamma',-\mathbf{k}\downarrow}^{\phantom{}} & c_{X,\mathbf{k}\uparrow}^{\dagger} & c_{X,-\mathbf{k}\downarrow}^{\phantom{}} & c_{Y,\mathbf{k}\uparrow}^{\dagger} & c_{Y,-\mathbf{k}\downarrow}^{\phantom{}}\end{array}\right)$, where $c_{\Gamma,\mathbf{k}\sigma}$, $c_{\Gamma',\mathbf{k}\sigma}$,
$c_{X,\mathbf{k}\sigma}$, and $c_{Y,\mathbf{k}\sigma}$ correspond
to operators on the $\Gamma$ and $\Gamma'$ hole pockets, on the
$X$ electron pocket at $\left(\pi,0\right)$ , and on the $Y$ electron
pocket at $\left(0,\pi\right)$, respectively. We have:

\begin{equation}
S=\int_{k}\Psi_{k}^{\dagger}\left(\hat{\varepsilon}_{\mathbf{k}}-i\omega_{n}\hat{1}\right)\Psi_{k}+\sum_{i=1}^{5}\int_{k}\chi_{i}^{-1}\left(k\right)\boldsymbol{\Phi}_{i,k}\cdot\boldsymbol{\Phi}_{i,-k}+\sum_{i=1}^{5}g_{i}\int_{k,q}\boldsymbol{\Phi}_{i,-k-q}\cdot\left(\Psi_{k}^{\dagger}\hat{\boldsymbol{\rho}}_{i}\Psi_{q}\right)\label{S_action}
\end{equation}
 where $k=\left(\mathbf{k},\omega_{n}\right)$ refers to both momentum
and fermionic Matsubara frequency, $\boldsymbol{\Phi}_{i,k}$ denotes
the collective bosonic fields associated with the SDW ($i=1,...,4$)
and Neel ($i=5$) fluctuations, and $ $$\chi_{i}\left(k\right)$
refers to the corresponding dynamic magnetic susceptibilities. Our
indices are defined such that $\chi_{\left(\Gamma,\Gamma'\right)}\left(\pi,0\right)$
corresponds to $i=1,2$, $\chi_{\left(\Gamma,\Gamma'\right)}\left(0,\pi\right)$
to $i=3,4$, and $\chi\left(\pi,\pi\right)$ to $i=5$. The coupling
constants satisfy $g_{i=1...4}=g_{\mathrm{SDW}}$ and $g_{5}=g_{\mathrm{Neel}}$.
We also have the band dispersions $\hat{\varepsilon}_{\mathbf{k}}=\mathrm{diag}_{4}\left(\varepsilon_{\Gamma,\mathbf{k}},\,\varepsilon_{\Gamma',\mathbf{k}},\,\varepsilon_{X,\mathbf{k}},\,\varepsilon_{Y,\mathbf{k}}\right)\otimes\tau_{3}$,
where $\tau_{i}$ are Pauli matrices in Nambu space. For a spin-rotationally
invariant system, and for the case of singlet pairing, we can focus
on the $z$-axis projection of $\hat{\boldsymbol{\rho}}_{i}$ \cite{S_Lonzarich},
given by:

\begin{eqnarray}
 &  & \hat{\rho}_{1}=\left(\begin{array}{cccc}
0 & 0 & \tau_{0} & 0\\
0 & 0 & 0 & 0\\
\tau_{0} & 0 & 0 & 0\\
0 & 0 & 0 & 0
\end{array}\right);\;\hat{\rho}_{2}=\left(\begin{array}{cccc}
0 & 0 & 0 & 0\\
0 & 0 & \tau_{0} & 0\\
0 & \tau_{0} & 0 & 0\\
0 & 0 & 0 & 0
\end{array}\right);\;\hat{\rho}_{3}=\left(\begin{array}{cccc}
0 & 0 & 0 & \tau_{0}\\
0 & 0 & 0 & 0\\
0 & 0 & 0 & 0\\
\tau_{0} & 0 & 0 & 0
\end{array}\right)\nonumber \\
 &  & \hat{\rho}_{4}=\left(\begin{array}{cccc}
0 & 0 & 0 & 0\\
0 & 0 & 0 & \tau_{0}\\
0 & 0 & 0 & 0\\
0 & \tau_{0} & 0 & 0
\end{array}\right);\;\hat{\rho}_{5}=\left(\begin{array}{cccc}
0 & 0 & 0 & 0\\
0 & 0 & 0 & 0\\
0 & 0 & 0 & \tau_{0}\\
0 & 0 & \tau_{0} & 0
\end{array}\right)\label{aux_matrix}
\end{eqnarray}

The Eliashberg equations are obtained by calculating the one-loop
self-energy 
\begin{equation}
\hat{\Sigma}_{k}=\sum_{i}g_{i}^{2}\int_{q}\chi_{i}\left(k-q\right)\hat{\rho}_{i}\hat{G}_{q}\hat{\rho}_{i}\label{Eliash1}
\end{equation}
 with $\hat{G}_{k}^{-1}=\hat{G}_{0,k}^{-1}-\hat{\Sigma}_{k}$ and
$\hat{G}_{0,k}^{-1}\equiv i\omega_{n}\hat{1}-\hat{\varepsilon}_{\mathbf{k}}$.

\subsection{Reformulation of Equations}

To solve the self-consistent system of equations, Eq.~\ref{Eliash1},
we rearrange them into a form more convenient for numerical solution.
We write the self energy as 
\begin{equation}
\hat{\Sigma}_{k}=i\omega_{n}\left(\hat{1}-\hat{Z}_{k}\right)\otimes\tau_{0}+\hat{W}_{k}\otimes\tau_{1}+\hat{\zeta}_{k}\otimes\tau_{3}\label{sigmaeliash}
\end{equation}
where $\hat{Z}_{k}=\mathrm{diag}_{4}\left(Z_{\Gamma,k},\, Z_{\Gamma',k},\, Z_{X,k},\, Z_{Y,k}\right)$
and $\hat{\zeta}_{k}=\mathrm{diag}_{4}\left(\zeta_{\Gamma,k},\,\zeta_{\Gamma',k},\,\zeta_{X,k},\,\zeta_{Y,k}\right)$
are the imaginary and real parts of the normal component, respectively,
and $\hat{W}_{k}=\mathrm{diag}_{4}\left(W_{\Gamma,k},\, W_{\Gamma',k},\, W_{X,k},\, W_{Y,k}\right)$
is the anomalous component of the self-energy. As we discussed in
the main text, the real part $\zeta_{\alpha,k}$ renormalizes each
$\alpha$ band dispersion (possibly in different ways \cite{S_Benfatto11}),
and will not be discussed here.

Hereafter we will consider two degenerate hole pockets ($N_{\Gamma}=N_{\Gamma'}$
and $\chi_{\Gamma}\left(\mathbf{Q}_{i}\right)=\chi_{\Gamma'}\left(\mathbf{Q}_{i}\right)$),
implying $W_{\Gamma}=W_{\Gamma'}$ and $Z_{\Gamma}=Z_{\Gamma'}$.
Using the tetragonal symmetry of the system, we have $\chi\left(0,\pi\right)=\chi\left(\pi,0\right)$
and $Z_{X}=Z_{Y}$, reducing the number of self-consistent equations
to five. We integrate over the momentum component $q_{\perp}$ perpendicular
to the Fermi surface, using the fact that the electronic propagator
is more sharply peaked at the Fermi level than the bosonic propagator
\cite{S_Chubukov}. Next, we linearize the equations by keeping the
leading terms in order $W_{\alpha,n}$ \cite{S_Carbotte} and average
the gaps along each Fermi pocket. We also include the impurity scattering
and the Coulomb pseudo-potential in the standard way, obtaining, for
the normal part:

\begin{eqnarray}
\left(Z_{\Gamma,n}-1\right)\omega_{n} & = & 2g_{\mathrm{SDW}}^{2}N_{X}T\sum_{m}\mathrm{sgn}\left(2m+1\right)\int dq_{\parallel}\chi_{\mathrm{SDW}}\left(q_{\parallel},\omega_{n}-\omega_{m}\right)\nonumber \\
 &  & +g_{u}^{2}\omega_{n}\sum_{\alpha}N_{\alpha}+u_{\mathrm{imp}}^{2}\mathrm{sgn}\left(\omega_{n}\right)\sum_{\alpha}N_{\alpha}\nonumber \\
\left(Z_{X,n}-1\right)\omega_{n} & = & T\sum_{m}\mathrm{sgn}\left(2m+1\right)\left[2g_{\mathrm{SDW}}^{2}N_{\Gamma}\int dq_{\parallel}\chi_{\mathrm{SDW}}\left(q_{\parallel},\omega_{n}-\omega_{m}\right)+g_{\mathrm{Neel}}^{2}N_{X}\int dq_{\parallel}\chi_{\mathrm{Neel}}\left(q_{\parallel},\omega_{n}-\omega_{m}\right)\right]\nonumber \\
 &  & +g_{u}^{2}\omega_{n}\sum_{\alpha}N_{\alpha}+u_{\mathrm{imp}}^{2}\mathrm{sgn}\left(\omega_{n}\right)\sum_{\alpha}N_{\alpha}\label{Z_eq}
\end{eqnarray}
 and for the anomalous part:

\begin{eqnarray}
W_{\Gamma,n} & = & -g_{\mathrm{SDW}}^{2}N_{X}T\sum_{m}\frac{\left(W_{X,m}+W_{Y,m}\right)}{Z_{X,m}\left|\omega_{m}\right|}\int dq_{\parallel}\chi_{\mathrm{SDW}}\left(q_{\parallel},\omega_{n}-\omega_{m}\right)\nonumber \\
 &  & -g_{u}^{2}T\sum_{m}\sum_{\alpha}N_{\alpha}\frac{W_{\alpha,m}}{Z_{\alpha,m}\left|\omega_{m}\right|}+u_{\mathrm{imp}}^{2}\sum_{\alpha}N_{\alpha}\frac{W_{\alpha,n}}{Z_{\alpha,n}\left|\omega_{n}\right|}\nonumber \\
W_{\left(X/Y\right),n} & = & -2g_{\mathrm{SDW}}^{2}N_{\Gamma}T\sum_{m}\frac{W_{\Gamma,m}}{Z_{\Gamma,m}\left|\omega_{m}\right|}\int dq_{\parallel}\chi_{\mathrm{SDW}}\left(q_{\parallel},\omega_{n}-\omega_{m}\right)-g_{\mathrm{Neel}}^{2}N_{X}T\sum_{m}\frac{W_{\left(Y/X\right),m}}{Z_{X,m}\left|\omega_{m}\right|}\int dq_{\parallel}\chi_{\mathrm{Neel}}\left(q_{\parallel},\omega_{n}-\omega_{m}\right)\nonumber \\
 &  & -g_{u}^{2}T\sum_{m}\sum_{\alpha}N_{\alpha}\frac{W_{\alpha,m}}{Z_{\alpha,m}\left|\omega_{m}\right|}+u_{\mathrm{imp}}^{2}\sum_{\alpha}N_{\alpha}\frac{W_{\alpha,n}}{Z_{\alpha,n}\left|\omega_{n}\right|}\label{W_eq_0}
\end{eqnarray}

Here, $g_{u}$ is the coupling to the Coulomb repulsion and $u_{\mathrm{imp}}^{2}$
is the averaged local impurity potential. The Coulomb repulsion renormalizes
all bare interactions, which become $g_{i}^{2}\rightarrow g_{i}^{2}/\left(1+g_{u}^{2}N\right)$,
where $N=2N_{\Gamma}+2N_{X}$ is the total density of states.

Using the diffusive expression for the spin susceptibility, Eq. (1)
of the main text, yields:

\begin{equation}
\int dq_{\parallel}\chi(q_{\parallel},\Omega_{n})\equiv\int_{-\infty}^{\infty}dq_{\parallel}\frac{1}{\left|\Omega_{n}\right|\gamma_{i}^{-1}+q_{\parallel}^{2}+\xi_{i}^{-2}}=\pi\xi_{i}\left(\frac{1}{\sqrt{1+\left|\Omega_{n}\right|\gamma_{i}^{-1}\xi_{i}^{2}}}\right)\label{chi_integrated}
\end{equation}

For convenience, we absorb the coefficient $\pi$ into the coupling
constants $g_{i}^{2}$ and introduce the density of states ratio $r=N_{X}/N_{\Gamma}$,
defining the Coulomb pseudo-potential $\mu^{*}=g_{u}^{2}N/\left(1+g_{u}^{2}N\right)$,
the scattering rate $\tau^{-1}=2N_{X}u_{\mathrm{imp}}^{2}$, and the
coupling constants $\lambda_{\mathrm{SDW}}=2g_{\mathrm{SDW}}^{2}\sqrt{N_{\Gamma}N_{X}}$
and $\lambda_{\mathrm{Neel}}=g_{\mathrm{Neel}}^{2}N_{X}$. An important
note about the form of the magnetic susceptibility: while its static
part comes from high-energy modes not considered in our model, the
Landau damping is a direct result of the coupling between the paramagnons
and the fermions, as given by Eq. (\ref{S_action}). Thus, the parameter
$\gamma_{i}$ contains information about the coupling constant $g_{i}$.
By evaluating the bosonic self-energy to one-loop, we obtain $\gamma_{\mathrm{SDW}}^{-1}=\kappa_{\mathrm{SDW}}^{-1}g_{\mathrm{SDW}}^{2}N_{X}N_{\Gamma}$
and $\gamma_{\mathrm{Neel}}^{-1}=\kappa_{\mathrm{Neel}}^{-1}g_{\mathrm{Neel}}^{2}N_{X}^{2}$,
where $\kappa_{i}$ are dimensionless parameters presumably of similar
orders of magnitude. This puts constraints on the ratio between the
effective couplings $\lambda_{\mathrm{SDW}}=2g_{\mathrm{SDW}}^{2}\sqrt{N_{\Gamma}N_{X}}$
and $\lambda_{\mathrm{Neel}}=g_{\mathrm{Neel}}^{2}N_{X}$, which is
expressed as $\frac{\lambda_{\mathrm{Neel}}}{\lambda_{\mathrm{SDW}}}=\frac{\kappa_{\mathrm{Neel}}}{2\kappa_{\mathrm{SDW}}}\,\frac{\gamma_{\mathrm{SDW}}}{\gamma_{\mathrm{Neel}}}\sqrt{\frac{N_{\Gamma}}{N_{X}}}$.

\subsection{Solution of Equations}

Substituting the definitions given above into Eq. (\ref{Z_eq}) and
evaluating the sums over Matsubara frequencies yields 
\begin{eqnarray}
\frac{Z_{\Gamma,n}\omega_{n}}{T} & = & \left(2n+1\right)+\sqrt{r}\lambda_{\mathrm{SDW}}\xi_{\mathrm{SDW}}S_{\mathrm{SDW},n}+\frac{\tau^{-1}}{T}\mathrm{sgn}\left(\omega_{n}\right)\left(1+\frac{1}{r}\right)\nonumber \\
\frac{Z_{X,n}\omega_{n}}{T} & = & \left(2n+1\right)+\left(\frac{\lambda_{\mathrm{SDW}}\xi_{\mathrm{SDW}}S_{\mathrm{SDW},n}}{\sqrt{r}}+\lambda_{\mathrm{Neel}}\xi_{\mathrm{Neel}}S_{\mathrm{Neel},n}\right)+\frac{\tau^{-1}}{T}\mathrm{sgn}\left(\omega_{n}\right)\left(1+\frac{1}{r}\right)\label{Z_finals-1}
\end{eqnarray}
 with the auxiliary functions:

\begin{eqnarray}
S_{i,n} & = & \frac{2\,\mathrm{sgn}\left(n\right)}{\sqrt{\frac{2\pi T\xi_{i}^{2}}{\gamma_{i}}}}\left[\mathrm{Hw}\left(\frac{1}{2},1+\frac{\gamma_{i}}{2\pi T\xi_{i}^{2}}\right)-\mathrm{Hw}\left(\frac{1}{2},\left|n\right|+\frac{\mathrm{sgn}\left(n\right)+1}{2}+\frac{\gamma_{i}}{2\pi T\xi_{i}^{2}}\right)\right]+\mathrm{sgn}\left(n\right)\,,\,\: n\neq0,-1\nonumber \\
S_{i,n} & = & 2\,\mathrm{sgn}\left(n\right)+1\,,\,\: n=0,-1\label{aux_Z_finals-1}
\end{eqnarray}

Here, $\mathrm{Hw}\left(\frac{1}{2},x\right)$ is the Hurwitz zeta
function for which efficient numerical evaluations exist.

After defining the auxiliary ``gap functions'' $\bar{\Delta}_{\Gamma,n}\equiv W_{\Gamma,n}/\left(\sqrt{N_{X}}Z_{\Gamma,n}\left|\omega_{n}\right|\right)$
and $\bar{\Delta}_{(X/Y),n}\equiv W_{(X/Y),n}/\left(\sqrt{N_{\Gamma}}Z_{X,n}\left|\omega_{n}\right|\right)$
and using the solutions for $Z$ given above, it is straightforward
to write down the gap equations (\ref{W_eq_0}) as a matrix equation
in Matsubara and band spaces, yielding Eq. (2) of the main text. For
numerical computations, it is convenient to use the tetragonal symmetry
of the system and split the matrix equation in two: one for the s-wave
case $\bar{\Delta}_{X,n}=\bar{\Delta}_{Y,n}$ and another one for
the d-wave case $\bar{\Delta}_{X,n}=-\bar{\Delta}_{Y,n}$, yielding
two different kernels $ $$\hat{K}_{s}$ ($2\times2$ matrix) and
$\hat{K}_{d}$ ($1\times1$ matrix). We use standard routines to obtain
the leading eigenvalue and corresponding eigenvector of the matrix;
the transition temperature is the temperature at which the leading
eigenvalue crosses zero.

\subsection{Equations in the orbital basis}

The formalism can be recast in a $N$-orbital basis in a straightforward
way. Consider the creation operator $a_{\mu,\mathbf{k}s}^{\dagger}$,
where $\mu=1,...,N$ refers to the orbital and $s$ to the spin. The
non-interacting Hamiltonian is given by:

\begin{equation}
H_{0}=\sum_{\mathbf{k}s}\sum_{\mu,\nu}\tilde{\varepsilon}_{\mu\nu,\mathbf{k}}a_{\mu,\mathbf{k}s}^{\dagger}a_{\nu,\mathbf{k}s}^{\phantom{\dagger}}\label{H0}
\end{equation}

By diagonalizing the matrix $\tilde{\varepsilon}_{\mu\nu,\mathbf{k}}$
in orbital space, $H_{0}$ can be written in terms of the band-basis
operators $c_{m,\mathbf{k}s}$:

\begin{equation}
H_{0}=\sum_{\mathbf{k}s}\sum_{m}\varepsilon_{m,\mathbf{k}}c_{m,\mathbf{k}s}^{\dagger}c_{m,\mathbf{k}s}^{\phantom{\dagger}}\label{H0_band}
\end{equation}
where $\varepsilon_{m,\mathbf{k}}=\sum_{\mu\nu}P_{m\mu,\mathbf{k}}\tilde{\varepsilon}_{\mu\nu,\mathbf{k}}P_{m\nu,\mathbf{k}}^{*}$
and $P_{\mathbf{k}}$ is the unitary matrix that diagonalizes Eq.
(\ref{H0}). It also follows that $c_{m,\mathbf{k}s}=\sum_{\nu}P_{m\mu,\mathbf{k}}a_{\mu,\mathbf{k}s}$.
Herefater, greek indices refer to orbitals and latin indices, to bands.

The magnetic susceptibility in the orbital basis is expressed as $\chi_{\mu\mu',\nu\nu'}$.
In principle, it can be calculated from the non-interacting susceptibility
$ $within an RPA approach, see Ref. \cite{S_Graser,S_Zhang}. By
defining the Nambu operators $\left(\begin{array}{cc}
a_{\mu,\mathbf{k}\uparrow}^{\dagger} & a_{\mu,-\mathbf{k}\downarrow}^{\phantom{\dagger}}\end{array}\right)$, and assuming a spin-rotationally invariant system with singlet pairing,
we obtain the one-loop self-energy:

\begin{equation}
\hat{\Sigma}_{\mu\nu,k}=\tilde{g}^{2}\int_{q}\chi_{\mu\mu',\nu\nu'}\left(k-q\right)\hat{G}_{\mu'\nu',q}\label{self_energ_orb}
\end{equation}
where $\tilde{g}$ is the coupling constant, $\hat{G}_{\mu\nu,k}^{-1}=\left(\hat{G}_{\mu\nu,k}^{(0)}\right)^{-1}-\hat{\Sigma}_{\mu\nu,k}$,
and the hat denotes a matrix in Nambu space. We have, in Nambu space:

\begin{equation}
\hat{G}_{\mu\nu,k}^{(0)}=\left(\begin{array}{cc}
G_{\mu\nu,k}^{(0)} & 0\\
0 & -G_{\mu\nu,-k}^{(0)}
\end{array}\right)\label{aux_G0}
\end{equation}
and:

\begin{equation}
\hat{\Sigma}_{\mu\nu,k}=\hat{\Sigma}_{\mu\nu,k}^{(N)}+\hat{W}_{\mu\nu,k}=\left(\begin{array}{cc}
\Sigma_{\mu\nu,k}^{(N)} & W_{\mu\nu,k}\\
W_{\mu\nu,k} & -\Sigma_{\mu\nu,-k}^{(N)}
\end{array}\right)\label{aux_sigma}
\end{equation}
where $\hat{\Sigma}^{(N)}$ denotes the normal part and $\hat{W}$,
the anomalous part of the self-energy. Defining the renormalized normal
Green's function $\mathcal{G}_{\mu\nu,k}^{-1}=\left(G_{\mu\nu,k}^{(0)}\right)^{-1}-\Sigma_{\mu\nu,k}^{(N)}$,
we write:

\begin{equation}
\hat{G}_{\mu\nu,k}=\hat{\mathcal{G}}_{\mu\nu,k}+\hat{\mathcal{F}}_{\mu\nu,k}=\left(\begin{array}{cc}
\mathcal{G}_{\mu\nu,k} & \mathcal{F}_{\mu\nu,k}\\
\mathcal{F}_{\mu\nu,k} & -\mathcal{G}_{\mu\nu,-k}
\end{array}\right)\label{aux_G}
\end{equation}
where $\mathcal{F}_{\mu\nu,k}$ is the anomalous Green's function.
Close to $T_{c}$, we can expand Dyson's equation to leading order
in $W_{\mu\nu,k}$, yielding:

\begin{equation}
\hat{G}_{\mu\nu,k}=\hat{\mathcal{G}}_{\mu\nu,k}+\hat{\mathcal{G}}_{\mu\mu',k}\hat{W}_{\mu'\nu',k}\hat{\mathcal{G}}_{\nu'\nu,k}\label{aux_G2}
\end{equation}

Substituting Eq. (\ref{aux_G}) gives the anomalous Green's function
in terms of $ $the anomalous self-energy:

\begin{equation}
\mathcal{F}_{\mu\nu,k}=-\mathcal{G}_{\mu\mu',k}W_{\mu'\nu',k}\mathcal{G}_{\nu'\nu,-k}\label{aux_F}
\end{equation}
which, combined with Eq. (\ref{self_energ_orb}), yields the Eliashberg
gap equation:

\begin{equation}
W_{\mu\nu,k}=-\tilde{g}^{2}\int_{q}\chi_{\mu\mu',\nu\nu'}\left(k-q\right)\mathcal{G}_{\mu'\mu'',q}W_{\mu''\nu'',q}\mathcal{G}_{\nu''\nu',-q}\label{Eliashberg_orb}
\end{equation}

It is now straightforward to make contact with our band-basis equations.
We can project on the band basis and obtain:

\begin{eqnarray}
W_{m,k} & = & -\tilde{g}^{2}\int_{q}\left[P_{m\mu,\mathbf{k}}^{*}P_{m\nu,-\mathbf{k}}^{*}\chi_{\mu\mu',\nu\nu'}\left(k-q\right)P_{n\mu',\mathbf{q}}P_{n\nu',-\mathbf{q}}\right]\mathcal{G}_{n,q}W_{n,q}\mathcal{G}_{n,-q}\nonumber \\
W_{m,k} & = & -\tilde{g}^{2}\int_{q}\left[P_{m\mu,\mathbf{k}}^{*}P_{m\nu,-\mathbf{k}}^{*}\chi_{\mu\mu',\nu\nu'}\left(k-q\right)P_{n\mu',\mathbf{q}}P_{n\nu',-\mathbf{q}}\right]\frac{W_{n,q}}{Z_{n,q}^{2}\omega_{n}^{2}+\varepsilon_{n,\mathbf{q}}^{2}}\label{final_Eliashberg}
\end{eqnarray}

The main difference between this expression and the one we used in
our calculations are the matrix elements $P_{m\mu,\mathbf{k}}$. As
pointed out by Refs. \cite{S_Graser,S_Maiti}, they can give rise
to an angular dependence of the gap function $W_{n,q}$ across Fermi
surface $n$ and maybe induce accidental nodes. 

However, in our Eliashberg formalism, we still average $W_{n,q}$
over the Fermi surface, since we are interested in comparing the energetics
of the $s$-wave and $d$-wave states. Thus, although these averaged
matrix elements will certainly renormalize the coupling constants,
they are not expected to change our main results. Furthermore, the
direct formulation of the problem in band space has the advantage
of allowing us to use the expressions for $\chi_{\mathrm{SDW}}\left(\mathbf{k},\omega\right)$
and $\chi_{\mathrm{Neel}}\left(\mathbf{k},\omega\right)$ as determined
by fittings of neutron scattering data.

\section{Form of the Real Axis spectral function}

The spectral function for Neel fluctuations that enters the Eliashberg
equations is 
\begin{equation}
A_{\mathrm{Neel}}\left(\omega\right)=\int dq_{\parallel}\mathrm{Im}\chi_{\mathrm{Neel}}\left(q_{\parallel},\omega\right)\label{Adef}
\end{equation}

Evaluating the integrals gives

\begin{equation}
A_{\mathrm{Neel}}\left(\omega\right)=\frac{\sqrt{2}\xi_{\mathrm{Neel}}}{\pi\sqrt{\gamma_{\mathrm{Neel}}}}\sqrt{\frac{\sqrt{\gamma_{\mathrm{Neel}}^{2}+\omega^{2}\xi_{\mathrm{Neel}}^{4}}-\gamma_{\mathrm{Neel}}}{\gamma_{\mathrm{Neel}}^{2}+\omega^{2}\xi_{\mathrm{Neel}}^{4}}}\label{A_omega}
\end{equation}

This is plotted for different $\xi_{\mathrm{Neel}}$ in inset of Fig.
2 of the main text.

\section{Bergmann-Rainer analysis}

We here recall how to obtain the functional derivative $\delta T_{c}^{\mathrm{s-wave}}/\delta A_{\mathrm{Neel}}\left(\omega\right)$
via the Bergmann-Rainer approach \cite{S_Bergmann_Rainer}. The linearized
gap equation for the s-wave channel can be cast as an eigenvalue problem
\begin{equation}
\eta\hat{\Delta}=\hat{K}_{s}\hat{\Delta}\label{GapEq1}
\end{equation}
 with $\hat{K}_{s}$ is a matrix in Matsubara and band space and $\eta$
the largest eigenvalue of $\hat{K}$. $T_{c}$ is the temperature
at which $\eta=0$.

Changing $A_{\mathrm{Neel}}(\omega)\rightarrow A_{\mathrm{Neel}}(\omega)+\delta A_{\mathrm{Neel}}(\omega)$
and using the Hellman-Feynman theorem gives 
\begin{equation}
k\frac{\delta\eta}{\delta A_{\mathrm{Neel}}(\omega)}=\left(\hat{\Delta}\right)^{\dagger}\frac{\delta{\hat{K}}_{s}}{\delta A_{\mathrm{Neel}}(\omega)}\hat{\Delta}\label{BR1}
\end{equation}

Using the result:

\begin{equation}
\xi_{i}a_{nm}^{(i)}=\int_{0}^{\infty}d\omega\frac{\omega A_{i}\left(\omega\right)}{\omega^{2}+\left|\omega_{n}-\omega_{m}\right|^{2}}\label{A_omega_def}
\end{equation}
 we obtain:

\begin{equation}
\frac{\delta\left(K_{nm}^{\alpha\beta}\right)_{s}}{\delta A_{\mathrm{Neel}}\left(\omega\right)}=-\frac{\lambda_{\mathrm{Neel}}}{\omega}\left(\begin{array}{cc}
0 & 0\\
0 & \delta_{nm}\mathrm{sgn}\left(\omega_{n}\right)f_{n}\left(\frac{\omega}{2\pi T}\right)+\frac{\omega^{2}}{\omega^{2}+\left|\omega_{n}-\omega_{m}\right|^{2}}
\end{array}\right)\label{deriv_K}
\end{equation}
 where $f_{n}\left(x\right)=x\,\mathrm{Im}\left[\psi\left(-n+ix\right)-\psi\left(n+1+ix\right)\right]$
and $\psi(x)$ is the digamma function.

Using the fact that $\eta$ varies smoothly with temperature and rearranging
the equation gives 
\begin{equation}
\frac{\omega\,\delta T_{c}}{T_{c}\,\delta A_{\mathrm{Neel}}\left(\omega\right)}=\left[\hat{\Delta}\frac{\omega\,\delta\hat{K}_{s}}{\delta A_{\mathrm{Neel}}\left(\omega\right)}\hat{\Delta}\right]_{\eta=0}/\left(-\frac{T\partial\eta}{\partial T}\right)_{\eta=0}\label{S_functional}
\end{equation}
 where the sum over Matsubara and band indices is left implicit. Here
we choose to consider the functional derivative with respect to $A(\omega)/\omega$
because at low frequencies this is the quantity which gives the scattering
rate, while if $A$ is concentrated at high frequencies the $1/\omega$
gives the usual BCS logarithm.

\begin{figure}
\begin{centering}
\includegraphics[angle=270,width=0.55\columnwidth]{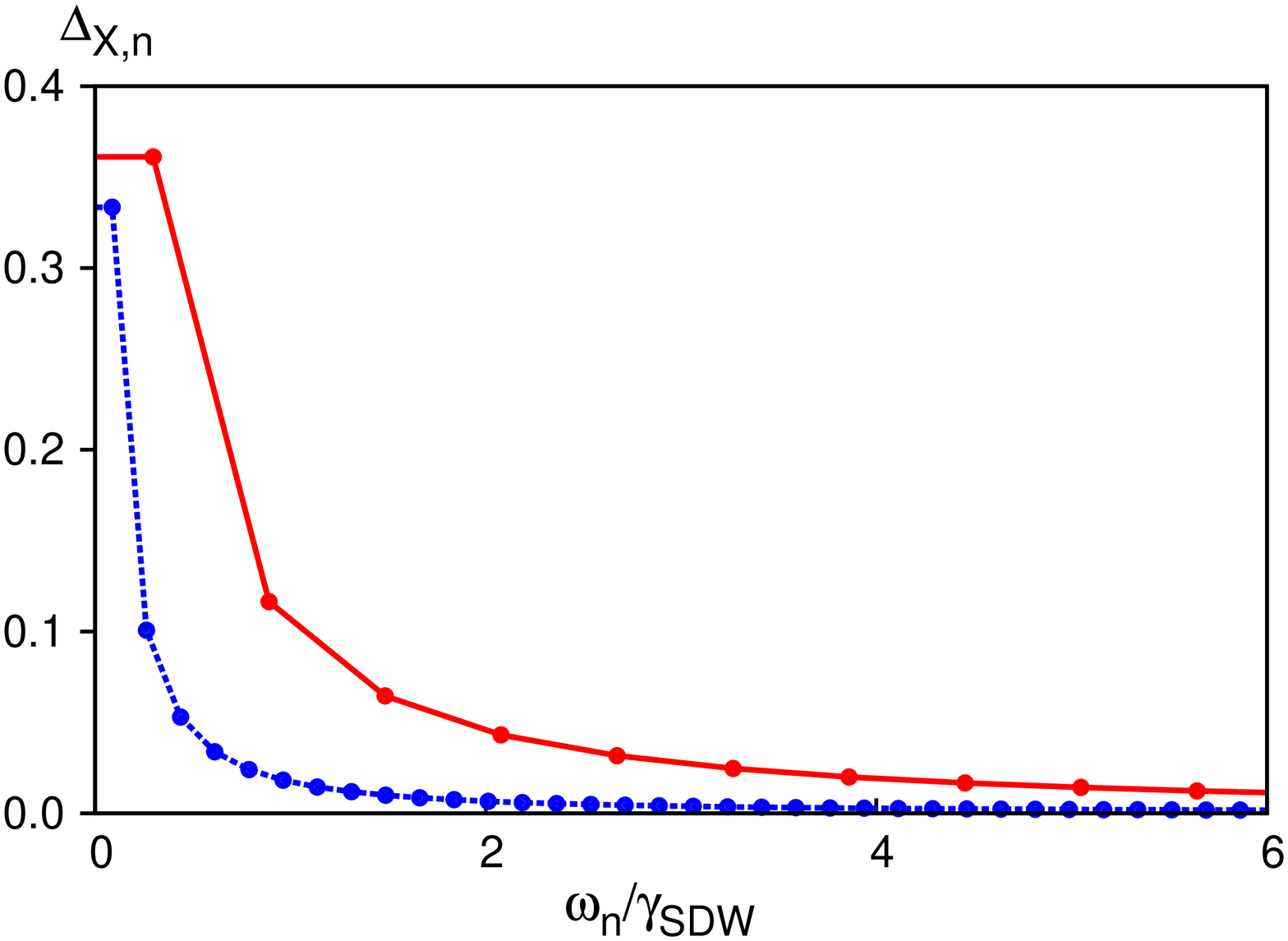} 
\par\end{centering}

\caption{Gap function of the electron pocket $\bar{\Delta}_{X,n}$ as function
of Matsubara frequency $\omega_{n}$ (in units of $\gamma_{\mathrm{SDW}}$)
for the parameters used in the main text and $\xi_{\mathrm{Neel}}=0$
(red/solid line) and $ $$\xi_{\mathrm{Neel}}=0.5a$ (blue/dashed
line). \label{fig_gaps}}
\end{figure}

We see that the logarithmic derivative of the transition temperature
(\ref{S_functional}) is determined by the frequency dependence of
the gap function (eigenvectors) $\bar{\Delta}_{i,n}$ -- explicitly
in the expectation value of $\delta\hat{K}_{s}$ and implicitly in
the temperature dependence of $\eta$. In figure \ref{fig_gaps},
we plot the gap $\bar{\Delta}_{X,n}$ of the electron pocket for the
cases $\xi_{\mathrm{Neel}}=0$ (red/solid line) and $\xi_{\mathrm{Neel}}=0.5a$
(blue/dashed line). The hole pocket gap displays a similar behavior.
Clearly, the $ $s-wave gap in the presence of Neel fluctuations decreases
much faster as function of frequency than the gap of the ``pure''
$s^{+-}$ state. Thus, the $s^{+-}$ state adapts to the presence
of $\left(\pi,\pi\right)$ fluctuations by suppressing the gap at
higher frequencies to avoid the main depairing effect, which comes
from the high frequency components of the Neel spectrum.

\section{Dependence of transition temperature on Impurity Scattering}

The Bergmann-Rainer approach is also used to obtain the dependence
of the transition temperature on impurity scattering: 
\begin{equation}
\frac{\partial T_{c}^{\mathrm{s-wave}}}{\partial\tau^{-1}}=\left[\hat{\Delta}\frac{\partial\hat{K}_{s}}{\partial\tau^{-1}}\hat{\Delta}\right]_{\eta=0}/\left(-\frac{\partial\eta}{\partial T}\right)_{\eta=0}\label{impurity}
\end{equation}
 where $\tau^{-1}$ is the scattering rate due to impurity scattering.
The reduced $\tilde{T}_{c}^{\mathrm{s-wave}}$ is then obtained via
the linear approximation $ $$\tilde{T}_{c}^{\mathrm{s-wave}}=T_{c}^{\mathrm{s-wave}}+\tau^{-1}\left(\frac{\partial T_{c}^{\mathrm{s-wave}}}{\partial\tau^{-1}}\right)$.

\end{widetext}

\end{document}